\documentstyle[ijmpel]{article}

\newcommand{\ps} { p\hspace{-0.18cm}/}
\newcommand{\Ps} { P\hspace{-0.22cm}/}
\newcommand{\bea}{\begin{eqnarray}}
\newcommand{\eea}{\end{eqnarray}}
\newcommand{\p}{\partial}
\newcommand{\g}{\gamma}
\newcommand{\be}{\begin{equation}}
\newcommand{\ee}{\end{equation}}
\newcommand{\nn}{\nonumber}
\begin{document}
\runninghead{Relativistic quantum kinetic equation$\ldots$} 
{Relativistic quantum kinetic equation$\ldots$}
\normalsize\textlineskip
\thispagestyle{empty}
\setcounter{page}{1}

\copyrightheading{}			

\vspace*{0.88truein}

\fpage{1}
\centerline{\bf RELATIVISTIC QUANTUM KINETIC EQUATION OF THE VLASOV TYPE}
\centerline{\bf FOR SYSTEMS WITH INTERNAL DEGREES OF FREEDOM}
\baselineskip=10pt
\vspace*{10pt}
\centerline{\footnotesize S.A. SMOLYANSKY and A.V. PROZORKEVICH}
\vspace*{0.015truein}
\centerline{\footnotesize\it Physics Department, Saratov State
 University, Saratov, Russia}
\baselineskip=10pt
\vspace*{10pt}
\centerline{\footnotesize S. SCHMIDT\footnote{permanent address: Fachbereich
Physik, Rostock University, Germany, \\email:
basti@darss.mpg.uni-rostock.de}\footnote{Work supported by Minerva foundation}}
\vspace*{0.015truein}
\centerline{\footnotesize\it School of Physics and Astronomy,
Raymond and Beverly Sackler Faculty}
\centerline{\footnotesize\it of Exact Sciences, Tel Aviv University, 69978  Tel
Aviv, Israel }
\baselineskip=10pt
\vspace*{10pt}
\centerline{\footnotesize D. BLASCHKE and G. R\"OPKE}
\vspace*{0.015truein}
\centerline{\footnotesize\it Fachbereich Physik, Rostock University, Germany }
\baselineskip=10pt
\vspace*{0.225truein}
\centerline{\footnotesize V.D. TONEEV }
\vspace*{0.015truein}
\centerline{\footnotesize\it Bogoliubov Laboratory of Theoretical Physics,
	Joint Institute}
\centerline{\footnotesize\it for Nuclear Research, 141980 Dubna, Russia  }
\baselineskip=10pt
\vspace*{0.225truein}


\abstract{ We present an approach to derive a relativistic kinetic
equation of the Vlasov type.  Our approach is especially reliable for
the description of quantum field systems with many internal degrees of
freedom. The method is based on the Heisenberg picture and leads to a
kinetic equation which fulfills the conservation laws. We apply the
approach to the standard Walecka Lagrangian and an effective chiral
Lagrangian.}

\vspace{1.5cm}


\newpage
\section{Introduction}

Experiments at CERN and BNL are designed to investigate the formation
and decay of a quark gluon plasma in ultra-relativistic heavy-ion
collisions\cite{QM}. The very early stage of such a collision is
dominated by non-equilibrium effects. The expected quark gluon plasma
undergoes a transition to strong interacting and highly correlated
hadronic matter if the system cools down and becomes dilute. The
description of such a dynamical evolution of a very complex,
relativistic system with many internal degrees of freedom is one of
the most challenging problems in modern particle and non-equilibrium
physics.

Relativistic kinetic theory and hydrodynamics have been successfully
employed for the description of non-equilibrium states of matter at
high density and temperature\cite{1,2,3}. These methods have been most
extensively applied in relativistic nuclear physics both in the
intermediate energy region\cite{4,5,6,7,8,9}, where the quark-gluon
degrees of freedom are still not relevant, and in the high energy
domain of a non-equilibrium quark-gluon plasma\cite{10}.  In the
construction of the relativistic kinetic theory, different approaches
have been used.  Within the contour Green function technique,
relativistic transport equations of the Vlasov type have been
developed by Kadanoff and Baym\cite{11}, Martin and Schwinger\cite{12}
and Keldysh\cite{13}.  That approach has been studied formally by a
large number of authors\cite{15}; and others have explored it in
quantitative model studies.

However, the theory can be developed further.  The derivation of a
relativistic transport equation that holds for systems with many
internal degrees of freedom as well as the inclusion of collision and
particle production source terms\cite{judah} are of particular
interest.

We investigate an approach based on the non-equilibrium statistical
operator in Heisenberg dynamics that can be considered as a
generalization of the Zubarev method\cite{24,25} to the relativistic
domain.  We restrict ourselves to the mean field approximation but
take into account the internal degrees of freedom of the constituents.

For demonstration we use two different models. The Walecka model as
the standard version of quantum hadrodynamics\cite{26} serves for the
description of hadronic degrees of freedom of relativistic nuclear
matter for laboratory energies less than a few GeV/A.  A further
application is a chiral Lagrangian that is of particular interest
because the concept of chiral symmetry breaking at large temperatures
and density is supposed to be the driving force for the hadronisation
of matter.  Therefore microscopic quark models, such as the
Nambu-Jona-Lasinio (NJL) model, may be useful for the dynamical
description of the expected quark-hadron phase transition\cite{ZW}.

The paper is organized as follows. In Section 2 we present an overview
of the method. In Sections 3 and 4 we demonstrate the derivation of
the relativistic kinetic equation of the Vlasov type using the Walecka
model and the NJL model, respectively.  We summarize the results in
Section 5.

\section{Equation of motion for the Wigner function}
In quantum field theory the equations of motion within the Heisenberg
picture are given as
\be\label{40}
i\partial ^\mu A(x)=[A(x),P^\mu ]\,\,,
\ee
where $A(x)$ is an arbitrary local operator and  $P^\mu $ is the
total  4-momentum of the  system,
\be\label{50}
P^\mu = \int d\sigma _\nu (n)T^{\mu \nu} .
\ee
Here $d\sigma ^\mu (n) $ denotes a vector element of an arbitrary space-like
hyperplane with a time-like normal vector $n^\mu $ $(n^2=1)$, $T^{\mu \nu
}(x)$ is the energy-momentum tensor. For the fermionic sector the
energy-momentum tensor $T^{\mu \nu }$ reads
\be\label{60}
 T^{\mu \nu
}=\frac i 4 \{{\bar \psi \gamma ^\nu \stackrel{\leftrightarrow}{\p ^\mu
}\psi + \bar \psi \gamma ^\mu \stackrel{\leftrightarrow}{\p ^\nu }\psi }\}-
g^{\mu \nu }{\cal L}(x).
\ee
The interaction is not specified yet and will be considered in the Walecka 
model (Section 3) and the NJL model (Section 4).

Let us now consider the situation when the time-like direction
in Minkowsky space is determined by an external
condition. We suppose in the following that this direction is defined by
the unit vector $n^\mu $ which fixes simultaneously the orientation of the
hyperplane $\sigma (n)$ \cite{31} in the formulae (\ref{50}) and
(\ref{20}). For the
description of the dynamics of the system along the time-like direction $n^\mu
$ and along the independent space-like directions on the hyperplane
$\sigma (n)$ , the boost transformation
with "velocity" $n^\mu $ can be used for the equations of motion (\ref{40}).
Technically,
it is achieved with the help of the
projection of Eqs. (\ref{40}) on the direction $n^\mu $ and on the hyperplane
$\sigma (n)$.  Convolution of (\ref{40}) with $n^\mu $ leads to the dynamical
equation
\be\label{70}
 i\, \frac {\partial A(x)}{\partial \tau }=[A(x),H(\tau )].
 \ee
  The parameter
\be\label{72}
 \tau =n_\mu x^\mu
\ee
plays the role of a proper time in a new coordinate
system.  The derivative along the direction
of $n^\mu $ in Eq.(\ref{70}) is given as
 \be\label{80}
 \frac {\partial }{\partial \tau }=n_\mu
 \frac {\partial }{\partial x_\mu } \quad .
\ee
$H(\tau )$ is the
 Hamiltonian of the system \cite{31,32},
\be\label{90}
 H(\tau )=n_\mu P^\mu=\int d\sigma
 (n)n_\mu T^{\mu \nu }n_\nu \, .
\ee
The remainder equations
\be\label{100}
 i\frac{\p}{\p x_\mu ^\perp} A(x)=[A(x),\Pi ^\mu (\tau )]
\ee
intend for a description
 of infinitesimal transfers of a system on the space-like hyperplane
$\sigma (n)$. The space-like derivative is defined by the relation
 \be\label{110}
\frac{\p}{\p x_\mu ^\perp}=\Delta ^\mu _\nu \frac{\p}{\p x_\nu},
\ee
 where
$\Delta ^{\mu \nu }=g^{\mu \nu }-n^\mu n^\nu $ is the projection operator,
$\Delta ^{\mu \nu }n_\nu =n_\mu \Delta ^{\mu \nu }=0$. And finally,
 \be\label{120}
 \Pi
 ^\mu (\tau)=\Delta ^\mu _\nu P^\nu =\int d\sigma (n)n_\lambda T^{\lambda
\nu }\Delta ^\mu _\nu
\ee
 is the space-like momentum vector of the system.
$H(\tau )$ and $\Pi ^\mu (\tau )$ are the generators of  local
infinitesimal normal and tangential diffeomorphismes to a hyperplane of a
constant proper time. Note that a  similar covariant decomposition of the 
generator of a motion group can be found  by the quantization of the Einstein
gravitation theory\cite{33}.

 In order to derive the drift integral of the relativistic kinetic equation it
is convenient to introduce the anti-commutation relations for fermionic fields:
\be\label{20}
\{ \psi (x),\bar \psi (x')\} = -iS(x-x'),\hspace{2cm}
\{ \psi (x),\psi (x')\} = \{ \bar \psi (x),\bar \psi (x')\}=0
\ee
for any points $x^\mu $ and $x'^\mu $   which belong to an arbitrary space-like
hyperplane $\sigma $~. Let us note the following property of the
commutation function of spinor fields\cite{1}
\be\label{30}
S(x)\mid _{x^0=0}=i\gamma ^0 \delta ^{(3)}(x)\,\,,
\ee
and
\be\label{150}
\int d\sigma (n|x)S_{\alpha \beta }(x-x')\,y(x')=in_\mu \gamma
^\mu _{\alpha \beta }y(x),\quad  x\in \sigma (n),
\ee
 where $y(x)$ is an arbitrary function of field operators and the spin is
denoted by the indices $\alpha$ and $\beta$.

The standard definition of
the one-particle  covariant Wigner function of the Fermi subsystem reads
\be\label{160}
f_{\alpha \beta }(x,p)=\int dy\, e^{-ipy}<P_{\alpha
\beta }(x,y)>,
\ee
where $<...>={\rm Tr}\{...\rho\} $ denotes the operation of
statistical averaging with the single particle density matrix $\rho$
\cite{24,25} in mean field approximation within the
Heisenberg representation with
\be\label{170}
P_{\alpha \beta }(x,y)=\bar \psi _\beta (x+y/2)\psi _\alpha (x-y/2)\,.
\ee
Performing the $\tau$ -
differentiation of the Wigner function (\ref{160}) and applying the
Liouville equation $d\rho /d\tau =0$   we find:
\be\label{180}
\frac {\partial f^{\alpha,\beta}(x,p)}{\partial \tau }=n^\mu\, \frac 
{\partial f^{\alpha,\beta}(x,p)}{\partial x^\mu}=\int dy\, e^{-ipy}<\frac 
{\partial }{\partial \tau }P^{\alpha,\beta}(x,y)>.  \ee The substitution of 
the equation of motion (\ref{70}) leads to the relation \be\label{190} 
n^\mu \frac{\partial}{\p x^\mu}f^{\alpha,\beta}(x,p)=-i\int dy
e^{-ipy}<[P^{\alpha,\beta}(x,y),H(\tau
)]>.
\ee
Let us assume that the  vector $p^\mu $ is time-like
and hence can fix the corresponding direction in the Minkowsky space, i.e.
\be\label{200}
 n^\mu \stackrel{def}{=}u^\mu =p^\mu /\sqrt{p^2},
\qquad u^2=1.
 \ee
Substitution of relation (\ref{200}) into Eq. (\ref{190}) leads to
\be\label{210}
 p_\mu \p ^\mu_x f^{\alpha,\beta}(x,p)=-i\sqrt{p^2}\int dy 
e^{-ipy} <[P^{\alpha,\beta}(x,y),H(\tau )]>.  \ee The right-hand side of 
 this equation is the Vlasov-like drift integral. This general result for 
the mean field approximation holds for different internal degrees of 
freedom (spin, isospin, color, flavor ...). In the following we will drop 
the spinor indices $\alpha,\beta$ in the notation.

\section{Application of the method to the Walecka model}
\subsection{General structure of the transport equation}
The simplest version of quantum hadrodynamics is the Walecka model
 of relativistic nuclear matter \cite{26}. The Lagrange
density for the nucleon ($\psi $), the neutral scalar ($\phi $)
and the vector ($\omega^\mu $) mesonic fields reads:
\be\label{10}
{\cal L}(x) =\frac{i}{2}\bar \psi \stackrel{\leftrightarrow}{\p}_\mu \! \g ^\mu
\psi - M\bar \psi \psi +
\bar \psi (g_s\phi -g_v\omega _\mu \gamma ^\mu )\psi +
 \frac 1 2[\partial _\mu \phi \partial ^\mu \phi-
m_s^2\phi ^2]-\frac 1 4 F^{\mu \nu }F_{\mu \nu }+
\frac 1 2 m_v^2\omega ^2 ,
\ee
where $M$, $m_s$, $m_v$  are the masses of
the nucleon, scalar and vector mesons, respectively; $g_s$, $g_v$ are
the coupling constants,
$F^{\mu \nu }=\partial ^\mu \omega ^\nu-\partial ^\nu \omega ^\mu $.
The meson fields are approximated by their mean fields, given by 
$\phi = <\phi>$
and $\omega_\mu = <\omega_\mu>$ where the symbol $<...>={\rm Tr}\{...\rho\}$
denotes the procedure of statistical averaging with the density matrix 
$\rho$ in the Heisenberg picture.

A concrete definition of the drift  integral is feasible only
by
specifying the Hamiltonian of the system (\ref{90}). In the considered
model, the Hamiltonian (\ref{130}) is a bilinear combination of the field
operators $\psi (x)$ and $\bar \psi (x)$, consequently the truncation
	 problem does not arise here.
For the Walecka model (\ref{10}) with the energy-momentum tensor (\ref{60}),
we obtain
\be\label{130}
H(\tau )= -\int d\sigma (n)\bar \psi \{\frac i 2 \gamma ^\mu
\stackrel{\leftrightarrow}{\p ^\perp _\mu }- M^* - g_v \gamma ^\mu
\omega _\mu \}\psi ,
\ee
\be\label{140}
\Pi ^\mu (\tau )=\frac i 4 \int
d\sigma (n)n_\lambda \bar \psi \{\gamma ^\nu \stackrel{\leftrightarrow}{\p
^\lambda }+\gamma ^\lambda \stackrel{\leftrightarrow}{\p ^\nu }\}\psi
 \Delta ^\mu _\nu ,
\ee
 where 
\[ M^*=M-g_s\phi \]
is the effective nucleon mass in mean field approximation.

 The intermediate calculations of the drift integral in the Walecka
model are shown in Appendix A. As it ought to be, the resulting relativistic
kinetic equation of the Vlasov type turns
out to be of non-local (and non-Markovian) type. The restriction
to minimal orders of the gradient expansions\cite{11} allows us to write
the relativistic Vlasov equation in a local form,
\bea\label{220}
 P_\mu\p^\mu_x f(x,P)+\frac 1 2 \p _\mu ^x M^*\{\Ps,\p
^\mu _P f(x,P)\} + g_vP^\mu F_{\mu \nu }\p ^\nu _Pf(x,P)+&&\nn\\\label{230}
\frac 1 2 [\Ps \g ^\mu ,\p _\mu ^x f(x,P)]+iM^*[\Ps,f(x,P)]-\frac 1 2
g_vF_{\mu \nu }[\Ps \g^\mu ,\p ^\nu _Pf(x,P)]&=&0 , \eea where
$P=p-g_v\omega$ is the kinetic momentum. The derivation of this kinetic
equation  is reviewed in\cite{20,23}.
Although the model (\ref{10}) is not gauge-invariant, this property emerges
when $m_v \rightarrow 0 $, therefore in the Walecka model it is
convenient to use the gauge-invariant generalization of the Wigner
function\cite{15,27,28}. In the case of small mesonic field
gradients this generalization can be done by the replacement
$p\rightarrow P=p-g_v\omega $ in (\ref{160}), this is fulfilled
in (\ref{220}).

The suggested method of the derivation of the relativistic kinetic equation of
the Vlasov type  does not lead to
the mass shell condition  itself. However  the space-like Eqs. of motion
(\ref{100}) generate also
a set of  equations for the Wigner function. Indeed, according to Eq.
(\ref{100}) the
result of the operation $\p _\perp ^\mu (x)$  (\ref{110}) on the Wigner
function (\ref{160}) leads to the equations\cite{23}
\be\label{240}
\frac{\p}{\p x_\mu^\perp }f(x,P)=-i\int dy e^{-ipy}<[P(x,y),\Pi ^\mu
(\tau )]>.
\ee
 The absence of an additional
restriction to the Wigner function is not obvious in general. However,
it is easy to show that these equations do not contain new
information in the mean field approximation of the Walecka model. 
The operator $\Pi ^\mu (\tau )$  is defined by relation (\ref{140})
which does not contain the mass and the mean field dependencies and,
consequently, the Eq.  (\ref{230}) can not result in the mass shell condition.
This conclusion is
confirmed by direct calculations of the right part of Eq. (\ref{230}) with the
operator (\ref{140})
\be\label{250} \frac{\p}{\p
x_\mu^\perp }f(x,P)+\frac 1 2 u_\nu [\g ^\mu \g ^\nu ,u_\lambda\p^\lambda_x
f(x,P)]+ i(p_\nu u^\nu)(2u^\mu f(x,P) - u_\nu \{\g ^\mu \g ^\nu
,f(x,P)\})=0 .  \ee 
One could assume that the mass shell condition is contained inside of the
kinetic equation (\ref{220}).
However it will be shown in the following section that this
assumption can't be confirmed.

\subsection{Dirac decomposition and properties of the relativistic Vlasov
equation}
In order to verify the physical consistency of the relativistic kinetic 
equation of the Vlasov type  (\ref{220}),
let us consider the simplest case of the spin saturated system
where the spin-dependent effects can be
neglected. Since the
pseudo-scalar and the pseudo-vector contributions in the model
(\ref{10}) are absent, the Wigner function decomposition on the basis of the
Clifford algebra is\cite{14,27}
\be \label{260}
f(x,P)=f^S(x,P)+f^V_\mu(x,P)\g ^\mu + f^T_{\mu \nu }(x,P)\sigma
^{\mu \nu } ,
\ee
where $\sigma ^{\mu \nu }=\frac i 2 \{\g ^\mu ,\g ^\nu\}$
and
\be\label{270}
f^S=\frac 1 4 {\rm Tr}\{f\} ,\qquad f^V_\mu =\frac 1 4 {\rm Tr}\{\g ^\mu f\}
,\qquad
f^T_{\mu \nu }=
\frac 1 8    {\rm Tr}\{\sigma ^{\mu \nu }f\}\,\,,
\ee
The symbol {\rm Tr} denotes the trace with respect to the spinor
indices.  Substitution of the relation (\ref{260}) into the kinetic equation
(\ref{220})
leads to the system of equations for the decomposition coefficients 
(\ref{270}).
Let us write this system in the quasi-classical limit $\hbar \rightarrow 0$
when  $f^T_{\mu \nu}=0$. Projection of the Dirac- scalar and  vector part of 
the Wigner function  leads to
\bea\label{280}
P_\mu\p^\mu_x f^S(x,P)+\p^x _\nu M^*P^\mu \p ^\nu _P f^V_\mu(x,P)+g_vP^\mu
F_{\mu \nu }\p ^\nu_P f^S(x,P)&=&0\\
P_\mu\p^\mu _x f^V_\lambda(x,P) + P_\lambda \p ^\mu _x f^V_\mu(x,P)
- P^\mu \p _\lambda ^xf^V_\mu(x,P) + \p _\nu ^x M^*P_\lambda \p ^\nu _P
f^S(x,P)&+&\nn\\\label{290} g_vP^\mu F_{\mu \nu}\p ^\nu _Pf^V_\lambda(x,P)
- g_vP_\lambda F^{\mu \nu}\p _\nu ^P f^V_\mu(x,P) + g_vP^\mu F_{\lambda \nu
}\p ^\nu _P f^V_\mu(x,P) &=& 0
\eea 
and
\be\label{300}
P_\mu f^V_\nu(x,P) =P_\nu f^V_\mu(x,P) \, .
\ee
For the description of a spin saturated system it is appropriate to find the
closed relativistic kinetic equation for the scalar part of the Wigner function
$f^S(x,P)$.

After a few algebraic transformations, we end up with the following kinetic
equation  for the scalar Wigner function:
\be\label{350}
P_\mu\p^\mu _x f^S(x,P)+\sqrt{P^2}(\p ^\mu _x M^*)\p _\mu ^P f^S(x,P)+g_vP^\mu
F_{\mu \nu }\p
^\nu _Pf^S(x,P)=0 .
\ee
Let us now discuss some feature of this relativistic Vlasov equation. Firstly,
the kinetic equation  (\ref{350}) occurs  without the mass shell condition: 
this condition can be introduced only from outside as an additional 
restriction. 
On the mass-shell
\be\label{360}
P^2={M^*}^2 ,
\ee
Eq. (\ref{350}) turns into the well-known relativistic kinetic equation of the
Vlasov type for spin saturated nuclear
matter\cite{5,6,7,15,14} 
\be\label{370}
 P_\mu\p^\mu_x f^S(x,P)+M^*\p ^\mu _x M^*\p _\mu ^Pf^S(x,P)+g_vP^\mu
F_{\mu \nu }\p ^\nu _Pf^S(x,P)=0 .
\ee
The derived kinetic equation fulfills the conversation laws. For example it is
easy to check with help of  (\ref{350}) that the baryon current
density
\be\label{380}
 j^\mu (x)=\int dP\, {\rm Tr}\{\g ^\mu f(x,P)\}=4\int dP\, f_V^\mu (x,P)=4\int
dP\frac {f^S(x,P)}{\sqrt{P^2}}P^\mu
\ee
 is fulfilled by the continuity equation
\be\label{390}
{\rm div}j(x)=0  \, .
\ee
The entropy conservation is verified  analogously and holds in the
investigated mean field approximation (no collisions).

\section{Application of the method to the Nambu--Jona-Lasinio-model}

In this section we want to derive a relativistic kinetic equation of the Vlasov
type for NJL model \cite{NJL}. The NJL model was successfully applied for the
description of chiral symmetry breaking \cite{NJL1}. The effective Lagrangian
reads
\be\label{510}
{\cal L}(x) = \frac{i}{2}{\bar \psi}(x)\partial_\mu\gamma^\mu\psi-m_0{\bar
\psi}(x)\psi(x)-G\bigg[({\bar \psi}(x)\psi(x))^2+ ({\bar
\psi}(x)i\gamma^5\psi(x))^2\bigg],
\ee
where $m_0$ is the current quark mass and G the coupling constant. Within this
model study we neglect color and flavor degrees of freedom. We restrict to the
mean field approximation (Hartree approximation) and obtain
\be\label{520}
{\cal L}^{\rm Hartree}(x) = \frac{i}{2}{\bar
\psi}(x)\partial_\mu\gamma^\mu\psi(x)-m_0{\bar \psi}(x)\psi(x)-\sigma(x){\bar
\psi}(x)\psi(x)- \pi(x){\bar \psi}(x)i\gamma^5\psi(x),
\ee
where the mean fields in the scalar and pseudoscalar channel are given as
\bea\label{530}
\sigma(x)&=&-G<{\bar \psi}(x)\psi(x)>\,\,,\\\label{540}
\pi(x)&=&-G<{\bar \psi}(x)i\gamma^5\psi(x)>\,\,.
\eea
The corresponding Hamilton density reads
\be\label{545}
H_{\rm int}(x)={\bar \psi}(x)\phi(x)\psi(x)\,\,,
\ee
with the channel decomposition
\be\label{550}
\phi_{\alpha\beta}(x)=
\sigma(x)\delta_{\alpha\beta}+\pi(x)i\gamma^5_{\alpha\beta}\,\,.
\ee
The substitution of Eq. (\ref{545}) into the kinetic equation  (\ref{210}) 
leads to the following equation for the model Lagrangian (\ref{520}) in lowest 
order of the gradient expansion
\bea
&&p_\mu\partial^\mu_xf(x,p)+
\frac{1}{2}[\ps\gamma^\mu,\partial_\mu^xf(x,p)]+im_0[\ps,f(x,p)]\nn\\
\label{560}
&+&i\{\ps\phi(x)f(x,p)-f(x,p)\phi(x)\ps\}+\frac{1}{2}\{\ps(\p_\mu^x\phi)
\p_p^\mu f(x,p)+(\p^\mu_pf(x,p)(\p_\mu^x\phi)\ps\}=0.
\eea
The kinetic part of this equation equals to the corresponding terms in Eq.
(\ref{220}). The last two terms are result of the interaction within the
NJL model.

In order to derive a closed equation for a spin saturated system we use the
Clifford decomposition of the Wigner function
\be\label{570}
f=f^s+i\gamma_5f^p+\gamma^\mu f^v_\mu+\gamma^\mu\gamma^5f^A_\mu\,\,.
\ee
In the quasiparticle approximation we can extract the following set of 
equations for the coefficients of the decomposition (Eq. (\ref{570}))
\bea\label{580}
p_\mu\partial^\mu_xf^s+(\partial_\mu^x\sigma)p^\nu\partial^\mu_pf_\nu^v&=&0\\
\label{590}
p_\mu\partial^\mu_xf^p-(\partial_\mu^x\pi)p^\nu\partial^\mu_pf_\nu^v&=&0\\
\label{600}
p_\mu\p^\mu_xf^v_\lambda+p_\lambda\partial^\mu_xf_\mu^v-p^\mu\p_
\lambda^xf_\mu^v+
p_\lambda\{(\partial_\mu^x\sigma)\partial^\mu_pf^s-
(\partial_\mu^x\pi)\partial^\mu_pf^p\}&=&0\\
\label{610}
M_\sigma f^p+\pi f^s&=&0\\
\label{620}
p_\mu f^v_\nu - p_\nu f^v_\mu&=&0\,\,,
\eea
where $M_\sigma = m_0 + \sigma$. In order to proceed we will restrict to the
chiral limit ($m_0 = 0$) and introduce the chiral invariant mass 
${\cal M}=\sqrt{\sigma^2+\pi^2}$ 
and the transformed Wigner function as 
${\cal F} = f^s/\sigma$. 
On the mass shell ($p^2 = {\cal M}^2$) we obtain for the scalar
part (${\cal F}$)
\be\label{630}
p_\mu\partial^\mu_x{\cal F}(x,p)+{\cal M}(x)\partial_\mu^x{\cal
M}(x)\partial^\mu_p{\cal F}(x,p)=0
\ee
using the additional relation $p\partial_x\sigma=p\partial_x\pi=0$.

The relativistic equation of the Vlasov type (\ref{630}) for the NJL model has
the same form as in \cite{neisel}. Starting from this equation it is very
interesting to study the dependence of the scalar Wigner function (quark
condensate) as a function of space and time\cite{Meshustin}. Therefore it is
obviously necessary to define reasonable initial conditions. Such a numerical
investigation is in progress but out of the scope of this work.

\section{Summary}
We have presented a derivation of a relativistic quantum kinetic
equation for a system with many internal degrees of freedom.  For
illustrative simplicity, in our analysis we employed a mean field
approximation.  Our approach is based on a non-equilibrium statistical
operator and can be considered as a relativistic generalization of the
Zubarev method.  We obtain a transport equation which fulfills the
conservation laws.

We applied the approach to the Walecka model and the NJL model.  In
particular we performed the spinor decomposition of the transport
equation for both cases.

In future studies the Vlasov equation should be improved by the
inclusion of collision integrals.  The solution of the relativistic
Vlasov equation for the NJL model is of particular interest for the
quark hadron phase transition, where a quark condensate should evolve
with time and space\cite{Meshustin}.  An interesting application of
the approach developed herein would be the numerical study of the
Vlasov equation in a microscopic quark model.

\nonumsection{Acknowledgment}

One of the authors (S.A.S.) is indebted to the Max-Planck-Gesellschaft
for the possibility to visit the Rostock University. He thanks Yu.S.
Gangnus and K. Morawetz for stimulating remarks. S.S. is grateful for
the financial support provided by Minerva foundation and thanks
V. G. Morozov and C.D. Roberts for helpful discussions.

\appendix{: Details of the derivation of the relativistic kinetic equation of
the Vlasov type}

In this appendix we add  some details of the derivation of
Eq.  (\ref{220}). The starting equation is the generalized kinetic equation
(\ref{210}). It is
convenient to write this equation in the form
 \be
 p\p (x)f(x,p)= - I(x,p),
\ee
 where the drift integral was introduced
\be
I(x,p) = i\sqrt{p^2}\int dye^{-ipy}<[P(x,y), H(\tau )]>.
\ee
In the considered model (\ref{10}), the Hamilton operator is defined by formula
(\ref{140}).
After calculation of the commutator in the relation (A.2) by the rules
(\ref{20}) and (\ref{160}), we get
\be
 I=I^{(m)}+I^{(k)}+I^{(f)} ,
\ee
where $I^{(m)}$ is the mass part of the drift integral,
\be
I^{(m)}_{\alpha \beta }(x,p)=iM[\ps,f(x,p)]_{\alpha \beta },
\ee
$ I^{(k)}$ is the kinetic part,
\be
 I^{(k)}_{\alpha \beta }(x,p)=\g ^\mu _{\g \delta }\int
dy e^{-ipy}<\bar \psi _\beta (x_+)\p ^\perp _\mu (x_-)\psi _\delta (x_-)
\ps _{\alpha \g }+
\p ^\perp _\mu (x_+)\bar \psi _\g (x_+)\psi _\alpha (x_-)\ps
_{\delta \beta }>\,\,,
\ee
and $I^{(f)}$  is the mean field part,
\bea
 I^{(f)}_{\alpha \beta}(x,p)&=&i\int dye^{-ipy}\times\nn\\
&\times&\{\ps _{\alpha \g }\Gamma _{\g \delta
}(x_-)<\bar \psi _\beta (x_+)\psi _\delta (x_-) - \ps _{\delta
\beta }\Gamma _{\g \delta }(x_+)<\bar \psi _\g (x_+) \psi _\alpha (x_-)>\},
\eea
and
\be
 \Gamma _{\alpha \beta
}(x)=-g_s\phi (x)\delta _{\alpha \beta }+g_v \omega _\mu (x)\g ^\mu
_{\alpha \beta }\,\,.
\ee
We use the short notation, $x_\pm = x\pm y/2$.
 The mass part (A.4) has been
written already with Wigner functions. In order to express the remaining parts
of the drift integral (A.3) in terms of the Wigner function it is necessary
to apply the  transformation to formulae (A.5) and (A.6).
In formula (A.5) let us consider the first term from the right part
\bea
&&\int dye^{-ipy}<\bar \psi _\beta (x_+)\p ^\perp _\mu (x_-)\psi _\delta
(x_-) >\ps  _{\alpha \g }\nn\\&=&                        \frac 1 2
\int dye^{-ipy}\{<\bar \psi _\beta (x_+)\p ^\perp _\mu (x)\psi _\delta
(x_-)+\bar \psi _\beta (x_+)\p ^\perp _\mu (x_-)\psi _\delta
 (x_-)> \}\ps _{\alpha \g }\nn\\
&=&  \frac 1 2 \int dye^{-ipy}\{<\bar \psi
_\beta (x_+)\p ^\perp _\mu (x)\psi _\delta (x_-)+
(\p ^\perp _\mu (x)\bar \psi _\beta (x_+))\psi _\delta (x_-)>\}
\ps_{\alpha \g }\nn\\
&=&
\frac {1}{2}  \p ^\perp _\mu (x)f_{\delta \beta }(x,p)\hat \ps _{\alpha
\g }.
\eea
Here  the following identity was taking into account
\[
 p^\perp _\mu =\Delta _{\mu \nu }p^\nu =0
\]
according to
definition  $\Delta _{\mu \nu } =g_{\mu \nu }-p_\mu p_\nu (p^2)^{-1}$. The
second term of (A.5) is transformed by analogy
\be
 \int dye^{-ipy}<(\p
^\perp _\mu (x_+)\bar \psi _\g (x_+))\psi _\alpha (x_-)> \ps_{\delta
\beta }=        \frac 1 2  \p ^\perp _\mu (x)f_{\alpha \g
}(x,p)\ps_{\delta \beta }.
\ee
As a result of substitution
of (A.8),(A.9) into (A.5) we get one contribution from the relativistic Vlasov
equation (\ref{220})
\be
I^{(k)}_{\alpha \beta }(x,p)=\frac 1 2 \p _\mu (x)[\ps \g ^\mu ,f(x,p)]_
{\alpha \beta }  .
\ee
The formulae (A.4) and (A.10) have the local form already. 
The mean field part (A.6) can be written in non-local form relatively to the 
Wigner function without effort. 
However, in order to write this fragment of the drift integral in local form, 
we perform a gradient expansion\cite{11}) of the meson fields (A.7) with 
respect to the rapid variable $y^\mu $,
\[
\Gamma_{\alpha\beta} (x_\pm)\simeq 
\Gamma_{\alpha\beta} (x)\pm 
\frac 1 2 y_\mu \p ^\mu (x)\Gamma_{\alpha\beta} (x)\,\,.
\]
The substitution of the decomposition into (A.6) leads to the result
\bea
I^{(f)}_{\alpha \beta }(x,p)\simeq \{\ps \Gamma (x)f(x,p)-f(x,p)\Gamma (x)
\ps \}_{\alpha \beta }
&+&      \frac 1 2 \{ \ps \p ^\mu (x)
\Gamma (x)\p _\mu (p)f(x,p)\nn\\&+&\p _\mu (p)f(x,p)\p ^\mu (x)\Gamma (x)\ps
\}_{\alpha \beta }.
\eea
Taking into account the definition (A.7), fulfilling the necessary
commutations and substituting the canonical momentum  $p^\mu $  by the
kinetic one $P^\mu =p^\mu -g_v\omega ^\mu $  , the relation
(A.11) can be rewritten in the form of the Eq. (\ref{220}).

\nonumsection{References}

\end{document}